\documentclass[twocolumn,preprintnumbers,amsmath,amssymb]{revtex4}

\usepackage{amsmath}    % need for subequations
\usepackage{graphicx}   % for figures
\usepackage{color}

\begin{document}

\title{The observer effect}
\author{Massimiliano Sassoli de Bianchi}
\affiliation{Laboratorio di Autoricerca di Base, 6914 Carona, Switzerland}\date{\today}
\email{autoricerca@gmail.com}   %optional

\begin{abstract}

Founding our analysis on the \emph{Geneva-Brussels} approach to the foundations of physics, we provide a clarification and classification of the key concept of observation. An entity can be observed with or without a scope. In the second case, the observation is a purely non-invasive discovery process; in the first case, it is a purely invasive process, which can involve either creation or destruction aspects. An entity can also be observed with or without a full control over the observational process. In the latter case, the observation can be described by a \emph{symmetry breaking} mechanism, through which a specific deterministic observational process is selected among a number of potential ones, as explained in Aerts' \emph{hidden measurement approach}. This is what is called a product test, or \emph{product observation}, whose consequences are that outcomes can only be predicted in probabilistic terms, as it is the case in typical quantum measurements. We also show that observations can be about \emph{intrinsic} (stable) properties of the observed entity, or about \emph{relational} (ephemeral) properties between the observer and observed entities; also, they can be about intermediate properties, neither purely classical, nor purely quantum. Our analysis allows us to propose a general conceptual characterization of quantum measurements, as observational processes involving three aspects: (1) product observations, (2) pure creation aspects and (3) ephemeral relational properties. We also discuss the important concept of \emph{non-spatiality} and emphasize some of the differences and similarities between quantum and classical/relativistic observations.


\end{abstract}

\maketitle

\section{Premise}
\label{premi}

The purpose of the present article is to provide a clarification and classification of the concept of observation, as used in physical sciences. To do so, we shall make extensive use of the findings of the \emph{Geneva-Brussels school} in the foundations of physics, which counts within its founders Josef Maria Jauch, Constantin Piron and Diederik Aerts (see for instance Refs.~\onlinecite{Piron1, Piron2, Piron3, Aerts1, Aerts2, Aerts3, Aerts4, Aerts5, Aerts6, Christiaens} and the references cited therein). Therefore, as a whole, the present essay can also be considered as a non-technical review of some of the results and intuitions developed in the last decades by this school, and more particularly by Diederik Aerts, in his \emph{creation-discovery view}~\cite{Aerts3,Aerts4} and \emph{hidden measurement approach}~\cite{Aerts4,Aerts7}.

However, in the present article we shall make some didactical choices that slightly differ from those employed by Aerts and collaborators. The most important one is about the concept of observation, which is the central theme of the present work. In the ambit of the creation-discovery view, it is usually stated that quantum measurements are not just observations, as they can provoke a real change of the state of the measured entity. In the following, we shall turn such a statement upside down  and adopt an opposite semantic view: observations are not only detections of what pre-existed in a physical system, but processes that in general can also provoke changes, and this remains true also outside of the microscopic domain. In other terms, our didactical point will be to show that observation is a much wider concept as is usually understood, which also includes the idea of transformation.  

Because of the nature of this essay, which on one hand goes over some (not sufficiently well) known results of the Geneva-Brussels school (and particularly those of Piron and Aerts) and on the other hand presents these same results with sometimes a slightly different perspective, it is clearly not very practical to always distinguish in the text the pieces of reasoning which are literally borrowed from Piron, Aerts and collaborators, and those instead that, although inspired by those reasoning, are  presented here in a slightly different form or perspective, without compromising the readability and flow of the text. 

Therefore, we encourage the reader who will be fascinated by the ideas presented in this work, to  go back to the original sources, to which we will often refer to in the text, to also appreciate the different expositional style, didactical choices, and pieces of explanations that were provided by their creators. 

But there are also -- we believe -- a few interesting original ideas and examples in our analysis, which as far as we can judge have not yet been made fully explicit by the above authors, like for instance our characterization of quantum measurements as observations of truly \emph{relational properties} between the observer and the observed entities, as a simple explanation for their ephemeral nature. 

That said, we might add that if we had had the chance to have access to the conceptual clarification presented in this article, when as young students we approached for the first time quantum physics, surely we would have found the content of this theory far less mysterious, and its conceptual understanding much less problematic.

\section{Introduction}
\label{Introduction}

Observation is a central concept in science and its correct understanding is evidently of the utmost importance. Outside of science, observation also plays a crucial role, as it is by observing that, generally speaking, we gather knowledge about reality, be it scientific or potentially scientific knowledge, according to nowadays classification.  

An important point we need to be aware of is that \emph{observation is not interpretation}. One thing is to observe and another thing is to draw conclusions from our observations; one thing is to \emph{describe} and another thing is to \emph{explain} the content of our descriptions. 

Generally speaking, we understand that observation should be a neutral activity, with no other goals than observation itself, whereas interpretation should be a much more scope-oriented activity; the (main) scope of interpretation in science being \emph{explanation} and, based on explanation, \emph{prediction}. 

Observation is typically associated to the experimental process of data collection, whereas interpretation is associated to the creation of explanatory models and theories. However, a radical distinction between observation and interpretation is not possible. There is indeed a lot of objectivity in our theoretical interpretations of the experimental data, considering that the evolution of our scientific theories is founded on a method of critical nature. 

But there is also a lot of subjectivity and conventionality (i.e., of interpretation) in our alleged objective observations. This is so because we necessarily see reality through the lenses of our theories, which determine  what, how, where and when to observe. To quote Mark Twain, ``For whoever has only a hammer, sooner or later everything else will seem like a nail.'' The hammer, in Twain's metaphor, represents our cognitive filters, our theories of reality, our paradigms and world-visions, that tell us what we are supposed to observe and how to do it.  

But an even more subtle point is to become aware that not only our observations are strongly biased by our theorizations and, vice versa, our theorizations are strongly biased by our observations, but also our comprehension of reality is strongly biased by our understanding of what an observation is, or is meant to be. 

The above remark is of course crucial in all ambits of human activity. In the present article however, we shall mostly limit our considerations  to physics, drawing on what are the consequences of a deeper analysis of the key concept of observation in order to clarify some of the important interpretational problems of quantum and classical/relativistic physical theories.

\section{Observation as an act of discovery}
\label{Observation as an act of discovery}

If we open a good dictionary, we can typically read that an observation is a process through which an observer can gain (and register) information about objects belonging to his/her reality. To fix ideas, let us give an example of a very simple act of observation. Imagine yourself in a forest. Your eyes are wide open and you simply look at the trees surrounding you. In other terms, using your eyes and  brain as an observational instrument, you detect the sunlight reflected by the trees, and by doing so you gather information about some of their properties, like for instance their spatial locations, dimensions, variety, colors, and so on. A crucial point in this observational activity is that it is completely \emph{non-invasive} with respect to the observed entities. You observe the trees but your observation has no effect on them. 

It is probably from observational examples of this sort, which are typical of the interaction of  human beings with their natural environment, that a sort of prejudice emerged, that we have the tendency to believe in an almost unconscious way: \emph{that it is always possible to observe the countless entities populating our reality without disturbing them, i.e., without influencing their state and evolution}~\footnote{To be truthful, we must say that such a prejudice strongly holds only for unanimated entities. Indeed, the observation of living entities, like when a hunter hides to observe a prey from afar, can possibly involve some very subtle levels of inevitable disturbance that could influence the behavior of the living entity being observed. The reality of such phenomena, like the so-called \emph{sense of being stared at}, is very controversial in scientific circles, despite some available evidences in their favor (see for instance Ref.~\onlinecite{JOCS}, a special issue of the \emph{Journal of Consciousness Studies} dedicated to this controversial subject). But independently of the truth of these phenomena, we must observe, for intellectual honesty, that the common belief within humans is that living entities are capable to sense, in some mysterious way, if other living entities are focusing their attention upon them.}.

The reason for the development of such a prejudice is quite obvious. We live in a terrestrial environment that is almost constantly illuminated by the light of our sun or the indirect light of our moon (and in more recent times by the artificial light of our appliances). Therefore, the entities populating our macroscopic reality are constantly emitting light, be it the light they directly produce or the light they reflect. This is how we came to know these entities (of course, we are simplifying here, as not only the visual sense is involved in the discovery of our environment), so that we usually consider them as being in their undisturbed condition when they do actually emit direct or indirect light. Thus, we believe that observing them is about collecting something they spontaneously offer to us, as if they were constantly sending messages out to the world, informing it about their actual condition. 

To put it figuratively, it is as if the world was constantly talking to us, without us asking anything specific, like a person performing a monologue, speaking her/his thoughts aloud to whoever is willing to listen. And by doing so, by listening to the messages that are spontaneously emitted by the different entities populating our reality, we are able to discover many of their attributes and properties. So, we could say that our most basic and common understanding of the concept of observation is that to observe an entity is to \emph{discover} what an entity is, without affecting its ``isness'' in whatsoever way.

This way of understanding the concept of observation, as a pure act of discovery, is also deeply rooted in physics, in the formalism of classical mechanics, although in an invisible way. Indeed, since observation is believed to have no effects on what is being observed, as it is just an act of discovery of what is already present in a system, there is obviously no need to explicitly represent the observer in a physical theory. Therefore, classical theories describe the states, properties and the evolution of physical entities by assuming a priori that such states, properties and evolution would be the same, should they be observed or not observed (i.e., discovered or not discovered) by an observer (typically a human scientist with her/his experimental apparatus).

\section{Asking questions}
\label{Asking questions}

But what if reality's monologue suddenly ceases? Or what if we are tired to only passively listen to that monologue and want to start a conversation, an interactive dialogue? Imagine once more being in the forest, but this time during a moonless night. This means that you cannot anymore observe the trees, as they are no more emitting or reflecting a light that your human eyes are able to detect. If you are nevertheless intentioned to observe them, you have then to take a more active stance in your observational process. For instance, you can light a torch and use it to illuminate the surrounding trees, to observe their spatial positions.

When you light a torch, you enter the reality's scene and transform the monologue into a dialogue. The act of lightening a torch with the purpose of projecting its light in the environment, is asking the entities living in that environment specific questions, whose responses correspond to the light they will diffuse back to you, that you are then able to detect, i.e., to observe.

Now, as is well known, the pressure exerted by an electromagnetic source, although extremely feeble, is not null. Such a pressure (that depends on whether the light is absorbed or reflected) is certainly not sufficient to exert any appreciable influence on the macroscopic structure of massive trees, but the situation dramatically changes in case one would be interested in observing, and therefore illuminating, much smaller objects. 

That light is able to exert a pressure, and therefore disturb the objects it illuminates, was well-known also before the advent of quantum physics. However, it wasn't considered as a problem, as it was also believed that it was always possible, at least in principle, to arbitrarily reduce the intensity of the light source and diminish its effective disturbance, thus allowing one to observe whatever entity -- macroscopic or microscopic -- without sensibly affecting its condition (using for this a more sensitive detecting instrument). Coming back to our dialogical metaphor, the classical assumption is that it is always possible to not influence the reality's responses, by simply whispering our questions and opening wide our ears to listen to the whispers of its responses.

Heisenberg has been the first to seriously put into question the validity of such a prejudice, by analyzing (with the help of Bohr) the functioning of an hypothetical gamma microscope, used to observe the tiniest possible entities: electrons~\cite{Heisenberg}. We shall not enter here into the many subtleties of the analysis of Heisenberg, that allowed him to provide a first heuristic argument in favor of his famous uncertainty principle. But let us point out what could be a possible misunderstanding regarding the reasons that brought him to conclude that we must abandon the idea that it is possible to observe an electron without affecting its state. 

The usual argument (simplifying here to the extreme) goes as follows: because of the quantum nature of light, which can only transfer energy and momentum in small indivisible packets, called photons, it is impossible to arbitrarily lower the intensity of a light beam illuminating an electron, below a certain level, seeing that the lowest possible intensity corresponds to the transfer of a single photon, and that a single photon still carries with it a certain finite amount of energy and momentum, and will thus produce a non-negligible effect of disturbance on a tiny electron. 

This however is not entirely exact, as nothing really prevents us from arbitrarily lowering the intensity of a light beam, if we really want to. In fact, this can be done by lowering not only the number of photons that are potentially transferred per unit time, but also by lowering the frequency of these photons, i.e., the frequency of the light beam. So, even though the nature of light is quantum, one can still produce, at least in principle, a light beam of arbitrarily low intensity, that will cause a negligible disturbance even to a tiny electron. 

So, why Heisenberg, in his famous gedankenexperimente, insisted in only using high-frequency gamma photons? The reason is well known: he was not only interested in asking the electron a specific question, regarding its position in space, but was also determined in obtaining from it a \emph{precise answer}. And since the resolution of an optical instrument is proportional to the wavelength of the used radiation, to obtain such a precise answer he was forced to stick to gamma photons, carrying a potentially non-negligible disturbance over electrons. 

What Heisenberg thus realized is that when we pretend from reality a very accurate reply to our interrogatives, we are not anymore guaranteed that these interrogatives will be without consequences on what is being observed. In other terms, we are not anymore guaranteed in that case that our observations are reduced to non-invasive acts of pure discovery.

\section{Observation as an act of destruction}
\label{Observation as an act of destruction}

A common prejudice is to believe that the problem highlighted by Heisenberg with his analysis of the gamma microscope is only pertinent to the micro-world. In other terms, it is usually believed that the inevitable and irreducible disturbance of the observer on the observed (the so-called \emph{observer effect}) is only one of the many strangeness of the micro-world, but that nothing of the sort can truly happen in our everyday macroscopic reality. What the present author believes instead, is that the real strangeness is the fact that we had to wait for Heisenberg to realize that our observations, whether concerning microscopic or macroscopic entities, cannot be reduced to mere acts of discovery. Observations are much more than that. 

Indeed, one thing is to observe reality with  a priori no ideas of what we are going to observe, and another thing is to become much more scope oriented in our observational processes, deciding in advance what we want to observe and making sure that we will be able to gather some precise information about it. Then, observation cannot anymore be understood as a purely non-invasive discovery activity.

As a paradigmatic example, consider a small piece of wood~\footnote{The paradigmatic example of a piece of wood, that we shall further analyze later in the article, was firstly introduced by Aerts, in his PhD thesis~\cite{Aerts1}, as a means to illustrate the important concept of experimental incompatibility.} and imagine that you want to \emph{observe its burnability}, that is, its property of being a burnable piece of matter. When you take such a decision, you are not anymore just collecting information that is spontaneously offered to you by the wooden entity, passively listening to its monologue, but  are now dramatically entering the scene by asking a very specific question and, more importantly, pretending a very specific answer. 

Asking such a question, and being assured to receive from the piece of wood entity a definite answer, requires to act on it in a very specific way, which will depend on your adopted definition for the concept of \emph{burnability}. Burnability, as we all know, is the capacity of a body, in certain conditions, to combine with oxygen to produce heat. There are of course different possible ways (and levels of sophistication) to define the burnability property of an entity~\footnote{This freedom in considering different possible definitions for the properties that we associate to physical entities reveals that there is an important level of conventionality in our description of the world. This doesn't mean that physical properties are totally arbitrary or subjective, but that there is also an important human part in their construction.}. But for the sake of our discussion, it will be sufficient to adopt the following very simple definition: ``A physical entity is burnable if, when we put it in contact with the flame of a match, for $30$ seconds, such a contact will trigger a reaction that causes the disintegration of the entity.'' 

Having defined what we mean by burnability, we are now in a position to observe it in our piece of wood entity. Obviously, there is only one possible way to perform such an observation: one has to take the entity, put it in contact with the flame of a match for $30$ seconds, and check if this will produce its disintegration. If it is the case, then we can say that we have succeeded in observing its burnability, otherwise, that we have failed to observe it.

To summarize, if we want to observe something specific, and not simply uncritically collect what reality offers to us, then we have to become much more active in our observational processes and ask specific questions (in the present case the question is: ``Is the piece of wood entity burnable?''). To receive a response to our questions we must implement them in operational terms, i.e., perform specific experiments. And when we do so, we may receive from the entity under investigation a positive or a negative answer, according to whether the observation will be successful (confirmed) or unsuccessful (unconfirmed)~\footnote{Such a protocol, consisting in asking a question by specifying the experiment to be performed and the rule to be used to interpret the results of the experiment in terms of ``yes'' (successful) and ``no'' (unsuccessful) alternatives, is usually called a \emph{yes/no-question}, \emph{yes/no-experiment}, or \emph{experimental project} in the ambit of the Geneva-Brussels school.}. However, such a dialogue cannot anymore be considered as a non-invasive discovery activity, as it may have dramatic consequences for the entity under observation, which may deliver its answer as a swan singing its last beautiful song!

\section{Observation as an act of creation}
\label{Observation as an act of creation}

In the previous sections we have considered two opposite forms of observation. One purely non-invasive, like when we stare at the trees in a forest, corresponding to a pure act of discovery, and one totally invasive, like when we observe the burnability of a piece of wood and by doing so we destroy the property, together with the entity that possesses it. 

The invasive, destructive process of observation is associated, as we have seen, with our choice of taking a more active part in the reality's drama, by asking specific questions and pretending precise answers. However, the destruction of the property which is being observed is not necessarily the rule and the opposite may occur as well: it may happen that we create the very property we are meant to observe! This may appear as a more perplexing statement, but this is only so because we are not usually aware of the fact that our acts of creation can also be considered as acts of observations, and vice versa. 

To illustrate our purpose, let us consider an example taken from human interactions. Human beings can manifest different psychological states, characterized by specific properties. Let us consider for instance \emph{suspiciousness}. When a person is suspicious (i.e., when the property of suspiciousness is actual for that person) it will usually exhibit a certain number of behaviors, typical of that condition, like for instance the one of looking to other people with a certain insistence. Let us take such a characteristic behavior as a rough characterization of the property of suspiciousness. Imagine then that you enter into a place, say a tearoom, and want to observe the suspiciousness of the persons therein. Assume that you know from other sources that these persons are not, in that precise moment, manifesting any suspiciousness. However, when you start observing them \emph{very attentively}, to detect their possible suspiciousness, your behavior will appear suspicious to them, so that in turn they may possibly start looking at you with a certain insistence. Accordingly, you might conclude from your observation that some of the persons in the tearoom are manifesting suspiciousness.  

This example is interesting as it reveals that we are not always aware that in certain instances we may be the very creators of what we believe we are instead just discovering. A possible objection here could be that sentient beings may behave differently than inanimate entities, as we also admitted in the footnote of Section~\ref{Observation as an act of discovery}. To meet that objection, let us consider a different example. 

Consider a small solid object of whatever shape, made of a non-elastic material, and imagine that you want to observe its \emph{incompressibility}, which for the purpose of the present example we shall define as follows: ``A physical entity is incompressible if, when submitted to the action of a machine press exerting a pressure of $10,000$ pascal, it experiences a volume change of no more than $1\%$ of its initial volume.''

When we perform the above observation, i.e., when we submit the entity to the $10,000$ pascal of pressure in the press, the result can either be  successful or unsuccessful, according to the material the entity is made of and its shape. In the case the outcome of the observation is successful, which means that following the action of the press the entity's volume change is less than $1\%$, we can conclude that we have effectively observed the incompressibility of the entity. However, we certainly cannot affirm that the observational process has created the incompressibility, as the property was clearly already possessed by the entity before the observation was carried out. 

But what if we fail to observe the incompressibility, i.e., what if the entity's reduction of volume is greater than $1\%$? Then, in this case, we can conclude that prior to our observation the entity wasn't incompressible, as our observation confirmed. However, we must also conclude that following our observation the entity has acquired the property of being incompressible. Indeed, should we decide to perform again the test, then we know in advance, with certainty, that the outcome of the observation would be successful. This is because to observe the incompressibility property we had to compress the entity, and a (non-elastic) compressed entity is also an incompressible entity, according to our definition of incompressibility. In other terms, we must conclude that our observational process has created the very property it was meant to observe. 

This may appear a little bizarre, as by failing to observe the incompressibility, the effect of our observation is nevertheless the creation of incompressibility, but this is how things are, and we must thus conclude that an observation can very well create what is being observed, not only in the domain of psychological interactions between humans, but also in the domain of interactions between inanimate entities. 

\begin{figure}[!ht]
\centering
\includegraphics[scale =.7]{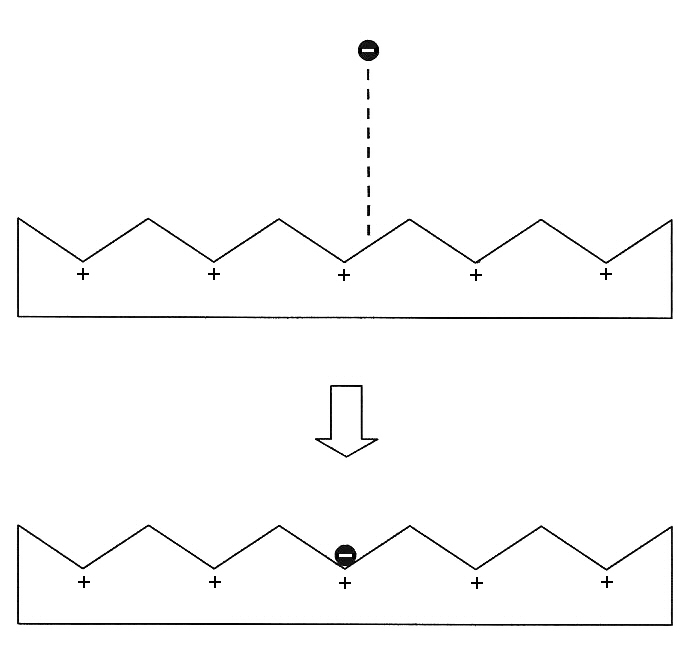}
\caption{Observing the horizontal position of a point particle using a sawtooth ruler.
\label{sawtooth ruler}}
\end{figure}

In fact, we can also easily create examples where the observation doesn't need to fail, to create the  property which is being observed. Consider for this a two-dimensional point-like classical particle of negative electric charge, and imagine you want to observe its \emph{horizontal spatial position}. Assume that scientists living in this unusual universe can only observe horizontal positions by employing strange \emph{sawtooth rulers} (see Fig.\ref{sawtooth ruler}), whose cavities are positively electrically charged. The observational procedure is as follows: ``Place the sawtooth ruler along the horizontal axis below the particle, then wait until the particle is drawn into one of the ruler's cavities, whose position will then give the measured horizontal position for the particle.''

Clearly, apart from the very special circumstances when the particle is already placed in perfect correspondence to the center of one of the ruler's cavities, the above observational procedure does literally create the horizontal position it observes (i.e., measures).

\section{Classifying observations considering their effects}
\label{Classify observations considering their effects}

So far, we have only considered the effects of a process of observation in relation to the specific property that is being observed, but of course an observation can potentially affect (create or destroy) not only the property which is under consideration, but also other properties that the observed entity possesses, and this can be the case also when the observation is of the discovery kind.

Let us come back to our piece of wood entity and imagine we want to observe its \emph{floatability}. Assume that we define floatability as follows: ``A physical entity floats if, when it is totally immersed into the water, it experiences an upward (buoyancy) force greater than its weight, causing it to reemerge from the water.'' 

Of course, a piece of wood entity, if made of normal (not too dense) wood, does possess the floatability property, which therefore can be successfully observed. Contrary to burnability, the observational process of the floatability property is not destructive for the entity. Indeed, at the end of the observational process we still have an integral piece of wood. Also, the observation of floatability doesn't destroy floatability, as is clear from the fact that, if one would decide to perform again the observation, the positive result would be once more certain in advance. However, we cannot say that the test is totally non-invasive for what concerns other properties possessed by the wooden entity, as is clear from the fact that its \emph{dryness} property is obviously (temporarily) destroyed by the observational procedure.  

Let us summarize our findings up to here. We have seen that an observation is a multifaceted process and that we can distinguish different categories of observation. Some observations involve a pure discovery aspect, with respect to what is being observed, whereas others involve pure creation or destruction aspects.

The observations involving a pure discovery aspect can be divided into two categories. We have the category of \emph{non-invasive discovery observations}, that leave the observed entity totally unaffected, like when we observe the trees in a forest by collecting the light they spontaneously offer to us. And we have the category of \emph{invasive discovery observations}, that leave the observed property unaffected, like when one observes the floatability of a piece of wood entity, without affecting it, but temporarily destroying other properties, like its dryness.

Then, we have the observational processes that involve pure aspects of creation or destruction of what is being observed. These are of course all invasive processes. An \emph{invasive creation observation} is a process that creates the property it observes (and possibly also affects other properties of the observed entity), like when we observe the suspiciousness of a human being, the incompressibility of a non-elastic object, or the horizontal position of a point-particle using a sawtooth ruler. On the other hand, an \emph{invasive destruction observation} is a process that destroys the property it observes (and possibly also affect other properties of the observed entity), like when we observe the burnability of a piece of wood.

\section{The classical prejudice}
\label{The classical prejudice}

In the previous section we have classified observations taking into account the possible effects they may produce on what is being observed. We want now to consider observations in relation to their \emph{predictability}. The question one can ask is if the outcome of a given observational process is in principle predictable with certainty, or not. According to the so-called \emph{classical prejudice}~\cite{Piron1,Piron2,Piron3}, the outcomes of all observational processes should be in principle predictable. As we shall show, such a prejudice is false, but before doing so, let us re-examine the different examples of observations we have considered in the previous sections, in the perspective of predictability. 

When we observe the trees in a forest, we can easily predict in advance their spatial locations, dimensions, variety and colors, if for example we  possess a detailed map of the place. Such a map is a description of the state of the forest, and it can easily be obtained from information gathered from previous observations. 

For what concerns the burnability and floatability properties of a piece of wood entity, again we can easily predict in advance the successful outcome of these observations. This is so because we have asked the same questions to countless wooden entities in the past, and have always received from them affirmative answers. Alternatively, we can infer these outcomes from a detailed analysis of the chemical and physical properties of wooden materials. But whatever is the method we employ to gather our knowledge about the wooden entity, the point is that we are perfectly able to predict in advance, with certainty, the effect of the burnability and floatability observational processes. 

The same we can also say, of course, for the observation of the incompressibility of a non-elastic material. Knowing in advance the characteristics of the material certainly allows us to predict if the observation will be successful or not. 

Also, our ability to predict with certainty the result of the observation of the position of a point-like particle, using a sawtooth ruler, only depends on our prior knowledge of the location of the particle, and more exactly on the precision with which we know its prior location. And since nothing prevents us from predetermining, with arbitrary precision, the spatial coordinates of a classical point-like particle, it surely is possible, at least in principle, to predict in advance the outcome of the sawtooth ruler observational process~\footnote{If the particle's horizontal position prior to the observation is exactly in correspondence with the tip of one of the teeth, then of course we cannot anymore predict with certainty the outcome of the observation, as it will be the result of a symmetry breaking mechanism.}. 

Finally, concerning the predictability of the outcome of the observation of suspiciousness, the situation is a little more controversial. Indeed, although a great majority of humans certainly respond in a very predictable way when they receive specific stimulis, it is also true that a minority of them sometimes react in a highly non-predictable way, independently of the inputs they receive from their environment. This possibility is sometimes called \emph{free will}. Now, if we assume that humans can, in certain circumstances, make use of their free will, then we cannot predict with absolute certainty, not even in principle, the outcome of observational processes performed on them (be they invasive or non-invasive), like for instance the observation of suspiciousness.

Apart from the fact that free will is a very controversial subject, as many thinkers believe it is just an illusion, one can object once more that living entities are intrinsically different in comparison to inanimate entities, and that the true mission of physics is to only study the latter, not the formers. Therefore, one may  ask if by limiting one's investigation to the observation of inanimate entities, one can conclude that, until proven to the contrary, it should always be possible, at least in principle, to predict the outcome of whatever observation, hence establishing the validity of the (deterministic) classical prejudice.

Of course, all quantum physicists know that quantum mechanics has invalidated such a prejudice, as nobody seems to be in a position to predict the outcomes of those observational processes called \emph{quantum measurements}. However, and contrary to what is usually believed, the falsity of the classical prejudice has nothing to do with the specificities of microscopic quantum systems, but with the way we generally attribute meet properties to physical entities, and the procedures we have to employ to observe them. 

To see this, we come back to the example of the piece of wood entity, which was firstly introduced and analyzed by Aerts as a simple illustration of a macroscopic physical system for which all the mystery of experimental incompatibility is under our eyes~\cite{Aerts1,Aerts11}. As we already emphasized, a (dry) piece of wood is a burnable entity. This means that if we decide to observe its burnability, we can predict with certainty, in advance, that the observation will be successful. And the same is clearly true for floatability.

Let us briefly open a  parenthesis to stress that every time we can in principle predict, with certainty, the successful outcome of the observation of a given property, without the need to execute in practice the observation, this is equivalent to affirm that such a property is an \emph{actual} property of the entity under consideration. On the other hand, if the prediction is not possible, not even in principle, then this is to say that the entity doesn't possess such a property or, more exactly, that it only possesses it in a \emph{potential} sense. Such a definition of the actuality and potentiality of a physical property has its origin in the famous  Einstein Podolsky and Rosen reality criterion~\cite{Einstein}, which was subsequently reworked by Constantin Piron~\cite{Piron1,Piron2,Piron3} to become a key ingredient of the Geneva-Brussels (realistic) approach to physics. (See also the discussion in Ref.~\cite{Massimiliano1}).

Now, as we said, the piece of wood entity does possess both properties: \emph{burnability} and \emph{floatability}. And this means that the piece of wood entity does also possess the \emph{meet property} of ``burnability \emph{and} floatability'' (the term ``meet'' is here used as in Boolean algebras), which is by definition the \emph{conjunction} of the burnability property \emph{and} floatability one. What we are simply stating here is a general fact about reality, namely that it is always possible to attribute several properties \emph{at once} to physical entities, and since we can do it, then also the property of possessing two or more properties at once must be a property of the system~\footnote{Regarding the question of attributing several properties at once, see in particular the interesting discussion in the introductory sections of Ref.~\onlinecite{Aerts11}.}. 

If what we are asserting here is correct, as it undoubtedly is, then we should also be in a position to define meaningful observational processes associated to \emph{meet properties}, and in our specific case to clarify what it means to successfully observe the ``burnability \emph{and} floatability'' meet property.

\section{How to conjunctly perform incompatible observations}
\label{How to conjunctly perform incompatible observations}

As the attentive reader has probably already realized, reading the article up to here, \emph{to observe a given property is to test it}. Indeed, when we successfully observe a property, we are doing nothing more than confirming (although not proving) its actuality. So, observing the ``burnability \emph{and} floatability'' meet property is about finding a single experimental procedure that would allow us to conjunctly test both properties. 

This may appear impossible at first sight. Indeed, the observation of floatability is destructive with respect to burnability, as is clear from the fact that a wet piece of wood doesn't burn anymore, according to our definition of burnability. In the same way, a burned piece of wood doesn't float any more, as ashes usually dissolve or precipitate when immersed in water. What we are saying is that the burnability and floatability observational processes are \emph{mutually incompatibles}. This should not surprise us more than necessary, if we just remember that we are speaking of invasive observational processes, that do affect the state and intrinsic properties of the observed entity. 

Let us not forget that the majority of our actions are invasive, and because of their invasiveness, they are generally incompatibles, in the sense that the effect of performing a given action $A$ (like putting on a sock) followed by another action $B$ (like putting on a shoe) is not in general the same as performing first the action $B$ followed by action $A$. Most of our observations being invasive processes, they will necessarily exhibit mutual incompatibilities, as it is the case of quantum observations (measurements), associated to non-commuting quantum observables. 

So, seeing that the observation of burnability is incompatible with the observation of floatability, how can we conjunctly observe them? In other terms, what does it mean to observe (or test) the ``burnability \emph{and} floatability'' meet property? The answer to such a tricky question was provided many years ago by Constantin Piron~\cite{Piron1,Piron2,Piron3}, by means of a very simple, and at the same time quite subtle, argument, which we are now going to explain. 

Considering that to observe is to test, what we simply need to find is a valuable test for the ``burnability \emph{and} floatability'' meet property, i.e., a test that would check, on a single piece of wood, the actuality of both properties (and therefore observe both at once). Such a test, which is usually called a \emph{product test} (or \emph{product question}, that we shall call here \emph{product observation}), consists in doing the following: ``First make a \emph{non-deterministic choice} regarding which one of the two observations you are going to perform, either the burnability or the floatability one, then perform it. If it is successful, you will say that also the product observation has been successful, i.e., that you succeeded in conjunctly observing the burnability and floatability properties.''

To understand why the above argument is correct, one must keep in mind what we have previously emphasized: that an entity does actually possess a property if and only if it is (in principle) possible to predict, with certainty, that its observation would yield a successful result, should we decide to perform it. The key point to understand is that we don't need to perform in practice the observation to infer the actuality of a property. It is only sufficient to be (in principle) in a position to predict with certainty the outcome of the observational process. 

Now, seeing that the product test procedure involves a non-deterministic choice, i.e., a choice whose outcome, by definition, cannot be predicted in advance by the observer, not even in principle, the only way s/he can guarantee that the observation would be successful, should it be performed, is to know in advance that its success is independent of such non-deterministic choice. And this can only be the case if the two properties -- the burnability and floatability properties in our case -- are both actual. 

So, although most of our observational processes are incompatible, as they correspond to invasive procedures, this doesn't mean we cannot define a meaningful procedure to observe (i.e., test) meet properties. But for this, we need to add to our observational processes a non-deterministic ingredient. 

For many readers the above reasoning may appear a little weird. This is because we are not used to think of our observations as experimental tests. But this is what, in ultimate analysis, goal oriented observations are all about: a way to check the existence (i.e., the actuality) of certain elements of reality, usually called properties (by discovering, creating or destroying them). And since in our reality entities can have more than a single property at once (this is how we construct reality!), it would be in fact much more weird if there wouldn't exist any possibility to test (i.e., to observe) such a state of affair~\footnote{This doesn't mean however that all meaningful properties of a physical system should necessarily be testable. See for instance the discussion in Ref.~\onlinecite{Smets}.}.

\section{The non-deterministic choice theorem}
\label{The non-deterministic choice theorem}

It is now time to come back to the classical prejudice and prove its falsity. More precisely, we have the following result.\\ 

\noindent\textbf{Theorem}: \emph{If meet properties are testable, then physical systems can behave non-deterministically.}\\ 

In other terms, the theorem affirms that if the fact that entities can possess many different properties at once is testable (i.e., observable), then the classical prejudice is false. To prove it, we must remember that to observe a property formed by the conjunction of two or more properties, one needs to employ a \emph{product observation}, which by definition requires a non-deterministic act of choice, i.e., a choice that the observer cannot predict in advance. So, if we assume that meet properties are testable properties, we must also assume that observers are able to perform purely non-deterministic choices. To show that this implies that physical systems, under certain conditions, can behave non-deterministically, let us consider a specific example of a property whose observation cannot be predicted in advance, not even in principle. 

For this, consider the \emph{non-burnability} property, which we define by means of the same observational procedure as for the burnability one, with the only difference that an unsuccessful outcome for the latter corresponds to a successful outcome for the former, and vice versa. Then, let us consider the ``non-burnability \emph{and} floatability'' meet property, i.e., the property of an entity to both possess the non-burnability and floatability properties, and ask the following question: ``Does the piece of wood entity possess such a meet property?'' 

Clearly, we cannot predict the outcome of the product observation associated to that question, not even in principle. Indeed, if it is the floatability test which will be chosen, the outcome will be successful, whereas if it is the non-burnability test that will be chosen, the outcome will be unsuccessful. In other terms, the observer cannot predict in advance the behavior (i.e., the response) of the piece of wood system, when s/he observes the ``non-burnability \emph{and} floatability'' meet property (i.e., asks a product question). S/he cannot do it even in principle, as the random choice element is an integral part of the very definition of the observational procedure, and cannot be eliminated without affecting the nature of the observed property. And this means that, whenever a meet property of this kind is observed, a physical entity as simple as a piece of wood is be able to exhibit a non-deterministic behavior.

It is worth noting that the above theorem simply expresses the fact that there is a structural correspondence between the behavior of an observer and the behavior of the observed entity. This is so because if there exist entities (be them human or not) that can manifest non-deterministic behaviors, then these entities can be used to actually perform product observations, through which meet properties can be observed. When an observer does so, s/he transfers her/his non-deterministic action to the entity under investigation, which therefore can also manifest a non-deterministic behavior, at least whenever certain meet properties are considered, like the ``non-burnability \emph{and} floatability'' one, in the case of a piece of wood.

The fact that not every observation is a classical (deterministic) observation was firstly observed by Piron, who is the designer of the product test procedure~\cite{Piron1,Piron2,Piron3}. If we have here presented such a result in the form of a rather pompous ``non-deterministic choice theorem,'' it is because of a recent interest concerning the relationship between human ``free will'' and the ``free will'' of microscopic systems, as substantiated in a famous free will theorem recently proved by John Conway and Simon Kochen~\cite{Con1,Con2}, stating that if experimenters can perform free choices, then the same must hold true for certain physical systems, like those formed by spin-$1$ twin particles. 

It is interesting to note that Conway and Kochen proved their theorem assuming (among other things) that under certain conditions (spacelike separation) our choices cannot affect the outcomes of certain experiments. On the other hand, the above non-deterministic choice theorem is based on the elementary observation that, on the contrary, the outcomes of our experimental observations are strongly dependent on our choices.

\section{The problem of control}
\label{The problem of control}

According to the above non-deterministic choice theorem, if we assume that meet properties are observables (i.e., testable) then the classical prejudice must be false. Therefore, there exist observational processes whose outcomes cannot be predicted in advance, not even in principle. This means that, in addition to our previous classification of Section~\ref{Classify observations considering their effects} (where observations were divided according to their effects), we can also divide observations according to their predictability, i.e., according to the fact that their outcomes are in principle predictable in advance, or not.

We want to now further persue our general analysis of the observational processes, and elucidate what could be a possible mechanism at the origin of the unpredictability of certain physical systems, like quantum systems, whose behavior, apparently, can only be described in probabilistic terms. 

An important point we have so far clarified is that many of our observations involve aspects of transformation, which can either consist in the creation or the destruction of what is being observed, or of some other properties possessed by the entity under investigation. When we talk about transformations, one of the things that comes to mind is the issue of \emph{control}. Do we control our creation/destruction processes, or don't we? Having control is about having the knowledge and the power to exert a specific action in order to precisely produce the desired effect.

If we come back to our dialogical analogy, having control is asking a question in a way that we can perfectly predict what will be the reply. It is not only about pretending an accurate reply from our interlocutor (as when Heisenberg asked about the exact position of an electron, using a gamma microscope), but also about obtaining a pre-determined reply, with certainty.  

Up to now, in our discussion, we have implicitly assumed that the observer has always a full control of her/his actions. For instance, we have assumed that s/he is able to immerse the piece of wood in a water vessel without knocking over the vessel, when s/he observes its floatability. Or that s/he will not change her/his mind during the execution of a certain experiment, leaving the observation half done, and therefore without any meaningful outcome! Certainly, we can always make sure to exert a full control over our observational processes, for as long as they involve simple and robust interactions. This however cannot be the case when we ask sophisticated product questions, requiring the use of non-deterministic choices which, by definition, are beyond our control. 

Another possibility is that the observational process involves a \emph{hidden mechanism} which would be responsible for the selection of a specific outcome; a mechanism that the observer would be totally unable to control, for a question of lack of knowledge and power. It is important to emphasize that we are not talking here about a lack of knowledge concerning the state of the observed entity (as it is assumed in certain hidden variables theories), but about the interaction between the observer (or the observer's measuring apparatus) and the observed entity. 

In other terms, we are distinguishing here a situation of possible incomplete knowledge of the state of the entity from a more subtle situation of lack of knowledge about the actual interaction that arises between the observing and the observed entity (which can also be understood as a lack of control by the observer over some aspects of the procedure of observation).

Clearly, if because of some hidden mechanism the observer cannot fully control the interaction, s/he will not be in a position to predict with certainty the outcome of the observation, which therefore s/he will only be able to describe, at best, in probabilistic terms. Can this explain the appearance of (non-Kolmogorovian) probabilities in typical quantum mechanical measurements? This is certainly possible, as Diederik Aerts demonstrated in his \emph{hidden measurement approach}~\cite{Aerts4,Aerts7}, showing that to a given quantum non-deterministic observation one can always associate a collection of ``hidden'' \emph{deterministic} interactions (called hidden measurements by Aerts), and that when the observation is performed (on an entity in a given state), one of these hidden interactions is selected and actually takes place. In other terms, according to this view, quantum probabilities would find their origin in the observer's lack of knowledge (or control) about which one of these hidden (deterministic) interactions does effectively take place during the observational process. 

In fact, the hidden-measurement mechanism is much more general than that, as it can be used to describe any probabilistic situation, and not only the quantum one. This means that it gives, in a way, a full description for what concerns the type of probability structures we can encounter in our world, as was proven in Section~4 of~\cite{Aerts14}.

\section{The (spin) quantum machine}
\label{The (spin) quantum machine}

Particularly important in the analysis of Aerts has been the possibility of substantiating his explanatory framework by means of a number of explicit macroscopic models, which can exhibit classical, quantum and intermediate behaviors, by simply varying the observer's level of control over the mechanism which is responsible for the selection of a specific interaction. One of the most famous of his examples is the (spin) \emph{quantum machine}, that we are going to now briefly describe~\cite{Aerts3,Aerts4,Aerts13}.

The physical entity under observation is a simple point-particle, localized on the surface of a three-dimensional Euclidean sphere of diameter $L$, the different possible states of which being the different places the point-particle can occupy. The particularity of the model resides in the way observations are carried out. Indeed, to observe the state (i.e., the position) of the entity, the experimental protocol is to use a sticky \emph{uniform} elastic band, to be stripped along a given orientation $\rho$, between two diametrically opposed points, $p_-$ and $p_+$, on the sphere's surface. Each chosen orientation $\rho$ for the elastic (defining a specific couple of end-points on the sphere's surface) characterizes a different observation with two possible outcomes. 

\begin{figure}[!ht]
\centering
\includegraphics[scale =.7]{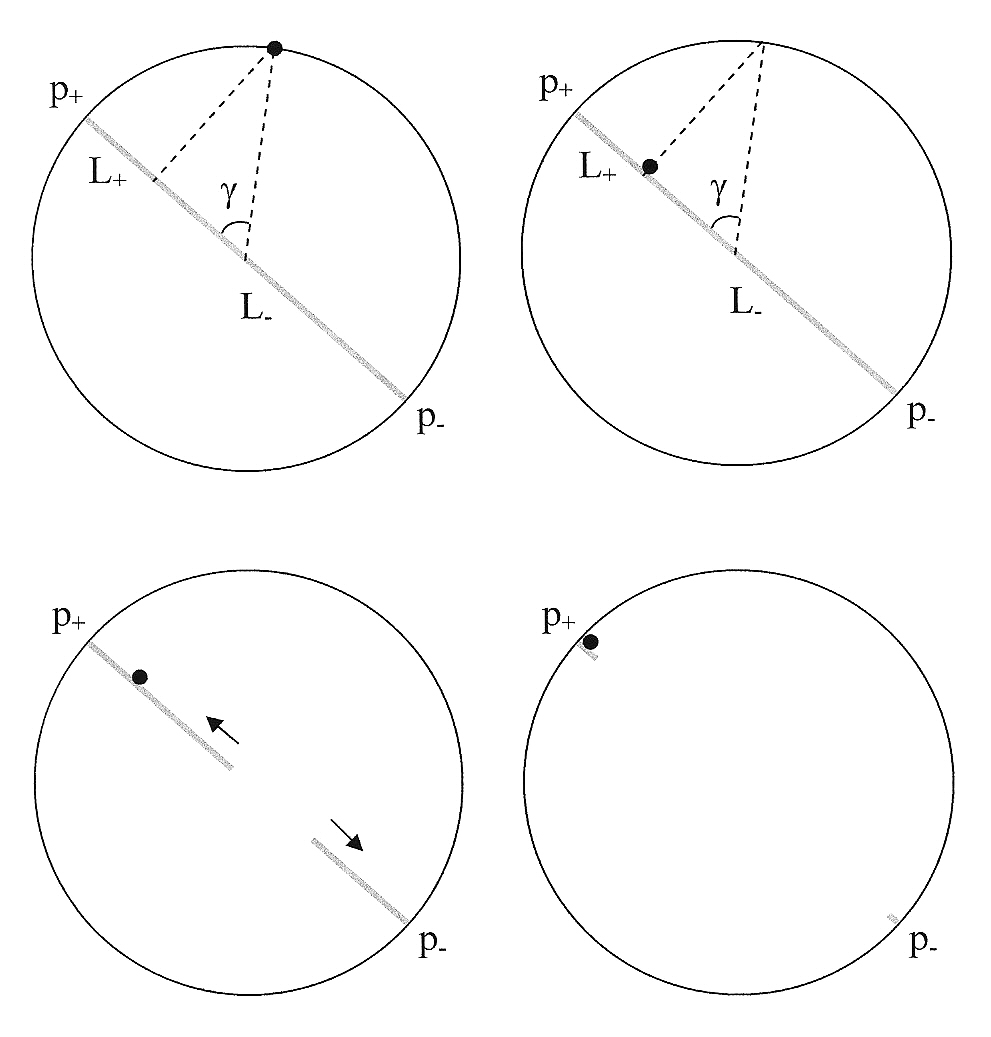}
\caption{A schematic representation of the quantum machine observational process, in the plane of the $3$-dimensional sphere where it takes place.
\label{spin quantum machine}}
\end{figure}

The experimental procedure is to let the point-particle fall from its original location orthogonally onto the elastic band, and stick to it. The emplacement of the particle stuck on the elastic thus defines two lengths, $L_\pm$, corresponding to the particle's distances to the end-points $p_\pm$. Then, one waits until the elastic breaks, at some unpredictable point, so that the particle, which is attached to one of the two pieces of it, will be pulled either toward $p_+$ or toward $p_+$, yielding in this way the specific outcome of the experiment, i.e., the spatial state which is acquired by the particle-entity as a result of the observational process (the different steps of the observation are schematically described in Figure~\ref{spin quantum machine}).  

With the help of some elementary trigonometry, it is easy to calculate the probabilities of the different possible outcomes. Indeed, the probability that the particle ends up in point $p_\pm$ is given by the length $L_\pm$ of the piece of elastic between the particle and the end-point, divided by the total length $L=L_+ + L_-$ of the elastic. Then, if $\gamma$ is the angle indicated in Figure~\ref{spin quantum machine}, between the direction of point $p_+$ with respect to the origin of the sphere and the direction associated to the particle's initial position, we have that the probabilities for the outcomes $p_\pm$ are given by:
\begin{equation}
\label{probabilities quantum machine}
{\cal P}_\pm = \frac{L_\pm}{L}=\frac{1}{2}(1\pm\cos\gamma)=
\begin{cases}
\cos^2\frac{\gamma}{2} \\
\sin^2\frac{\gamma}{2}. 
\end{cases}
 \end{equation}

The above are exactly the quantum probabilities one would obtain in a typical \emph{Stern-Gerlach measurement} with a spin-$1/2$ quantum entity, with $\rho$ corresponding to the chosen orientation of the Stern-Gerlach apparatus, defining the two possible directions for the spin  measurement, and $\gamma$ the angle between the apparatus' orientation and the direction of the prepared incoming spin state \cite{Aerts3,Aerts4,Aerts13}.

So, the observational process of Aerts' quantum machine is perfectly isomorphic to the observational process of a spin-$1/2$ quantum entity. This means that, thanks to his model, it is possible to make fully evident, in our conventional $3$-dimensional theatre, the hidden structure associated to the description of a two-dimensional quantum system, allowing to gain a better understanding about what really goes on (structurally speaking) when the quantum level of our reality interacts with our equipment.    

In fact, one can do much more, as it is possible to construct generalized quantum machines for arbitrary quantum mechanical entities~\cite{Aerts7,Aerts14,Coecke1,Coecke2,Coecke3}, meaning that the ``hidden measurements'' explanation can in fact be adopted to explain the origin of the quantum probabilities for general quantum systems~\footnote{Some readers may be tempted to believe that, because of the well-known Gleason's theorem~\cite{Gleason} and  Kochen-Specker's impossibility proof~\cite{Kochen}, it would be unfeasible to construct hidden-measurement models like the quantum machine one, for Hilbert spaces of dimension greater than 2. This is however not the case, as No-Go theorems for hidden variables only apply to models with hidden variables referring to the \emph{state of the system}, and not to models where the hidden variables refer instead to the \emph{observational process}; see in particular~\cite{Aerts-3d}, where an explicit model is constructed for a $3$-dimensional Hilbert space.}.

Aerts' quantum machine model also allows for the description of more general structures, beyond the usual classical and quantum ones. This can be done in the above example by choosing elastic bands with different characteristics. Indeed, if the elastics used are uniformly breakable (as we have assumed here above), then the observer is in a situation of maximum lack of knowledge about the point where the elastic will break (i.e., in a situation of maximum lack of control about which interaction is going to be actualized between the particle and the measuring apparatus), and as we have shown in this case the machine exactly reproduces the quantum probabilities. 

On the other hand, if the elastic used can only break in a single given predetermined point, then it becomes possible to predict in advance, with certainty, the outcome of the experiment, and we are in a situation of minimum lack of knowledge (maximum control of the observation), giving rise to a classical, purely deterministic description. 

But one can also consider intermediate cases, where the elastics used are uniformly breakable only on some of their segments. This corresponds to a generalization of the quantum machine, called the $\epsilon$\emph{-model}~\cite{Aerts3}, yielding more complex probability descriptions, neither classical nor quantum, but genuinely intermediate, corresponding to situations of intermediate knowledge, or control (more will be said about intermediate observations in Section~\ref{Intermediary observations}).

Aerts' quantum machine provides a simple and convincing mechanism explaining quantum probabilities as epistemic probabilities, originating from our lack of control/knowledge about what happens exactly during the observational process, i.e., during the interaction between the observer (also understood as the measuring apparatus) and the observed. More precisely, thanks to Aerts' machine, we can see that a quantum measurement is an observation that involves the selection of a specific observational process among a number of ``hidden'' observational processes (i.e., of ``hidden interactions''), which in the model correspond to the classical (invasive) observations (involving an aspect of creation) associated to the different points $x$ at which the uniform elastic band can break. 

More precisely, if $O_{u,\rho}$ is a given observation, characterized by a specific orientation $\rho$ of the uniform ($u$) elastic, we can associate it with a set of ``hidden'' observations $O_{x,\rho}$, $x\in [0,L]$, where $x$ specifies a possible breaking point of the elastic. In other terms, $O_{u,\rho}$ does not correspond to an \emph{actual} observation, but to a collection of \emph{potential} deterministic observations, one of which will be selected during the execution of the experimental procedure, in a way that is beyond the power control of the participative scientist. 

Therefore, we can say that a typical quantum observation (i.e., a quantum measurement) is a \emph{product observation}, involving a \emph{symmetry breaking} mechanism through which an \emph{actual} observation is selected among a number of \emph{potential} observations, in a way that cannot be controlled (and therefore predicted) by the observer. If we can speak here of a process of symmetry breaking it is because all the ``hidden'' potential observations have the same a priori chance to be actualized, as the elastic is assumed to be uniform.  

It is interesting to remark that if quantum measurements are product observations, this also means that they correspond to observations of \emph{meet properties}. In Aerts' quantum machine we have simply called such a meet property the \emph{position} of the point-particle, but to be more precise, we should have called it the \emph{position of the point-particle observed using a uniform elastic band stripped along orientation $\rho$}, which in short we can simply call the \emph{$(u,\rho)$-position} of the point-particle. This $(u,\rho)$-position can be understood as the meet property formed by the conjunction of \emph{$(x,\rho)$-position} properties, where a $(x,\rho)$-position is the \emph{position observed by means of an elastic band stripped along orientation $\rho$, that will break with certainty at point $x$}. So, the quantum observation of the $(u,\rho)$-position is a product of deterministic observations of $(x,\rho)$-positions, realized by means of elastics that can break only at specific points. 

Now, the interest of explicit models like the quantum machine is that, among other things, they allow us to see what is usually hidden. Indeed, in the model we can for instance easily replace the non-deterministic observation of the $(u,\rho)$-position by a deterministic observation of the $(x,\rho)$-position, by simply performing the experiment with an elastic which only breaks at point $x$, instead of a uniform one. This means that if we should want to, we could take full control of the observational process, and predict in advance its outcome. 
  
This however we cannot do (at least for the time being) when dealing with microscopic systems. When we observe for instance the spin of a spin-$1/2$ quantum entity, by means of a Stern-Gerlach apparatus, we know that, structurally speaking, we can understand it as a product observation, with a symmetry breaking mechanism operating ``behind the scenes.'' But we don't know what are exactly these hidden (potential) observations in the case of microscopic entities, that would be responsible for the emergence of the non-Kolmogorovian quantum probabilities. The hidden measurement approach just tells us what to search, but not what to find: we have to search not for hidden variables associated to the state of the entity, but for hidden variables associated to the observational process, i.e., to the ``pure'' (possibly deterministic) observational processes which are selected through a symmetry breaking mechanism that cannot be controlled by the participative scientist.

\section{Intrinsic VS Relational observations}
\label{Intrinsic VS Relational observations}

In the previous section we have described Aerts' quantum machine model and interpreted its observational processes as products of deterministic ``hidden'' observations. Although we don't know to what these ``hidden'' observations would correspond to, in the case of microscopic entities, Aerts' model has certainly the merit of lifting one corner of the veil of quantum mystery, demystifying part of quantum strangeness.

According to the \emph{non-deterministic choice theorem} of Section~\ref{The non-deterministic choice theorem}, we have seen that, strictly speaking, we must renounce the old classical prejudice, as it is always possible to create product questions such that their answers, by definition, cannot be predicted in advance. On the other hand, we have also seen that quantum measurements have the structure of product questions. This explains why their outcomes can only be predicted in probabilistic terms. But this also leaves the door open as to the possibility of their deconstruction in terms of deterministic sub-observational processes. 

So, if on one hand the analysis of the general structure of the questions we are allowed to address to reality force us to abandon the deterministic prejudice, on the other hand we are certainly not forced to abandon it because of quantum mechanics, as one can understand quantum observations as products of deterministic observations, implemented by a symmetry breaking mechanism.

This said, we are now going to analyze observations according to another important criterion: \emph{intrinsicness}. To introduce this concept, let us ask what would happen if, after some time, we would repeat the observation of the $(u,\rho)$-position of the point-particle in the quantum machine model. Assuming that in Aerts' toy universe the point-particle is a static entity, not moving by itself, then the repetition of the same observation $O_{u,\rho}$ would yield exactly the same outcome, because once a specific $(u,\rho)$-position is observed (created), it will clearly remain an \emph{intrinsic} stable property of the point-particle entity.

The situation is similar to those we have already considered in Section~\ref{Observation as an act of creation}, in relation to the observation of the \emph{incompressibility} of a non-elastic object, or of the \emph{horizontal position} of a point-particle as measured by a sawtooth ruler (assuming also in this case that the particle is motionless). In all these examples, once the property is observed, it can also be re-observed in the future, with certainty, should one perform again exactly the same observation. And this means that the property in question has become an \emph{intrinsic} property of the entity, i.e., a property that is stably actual for it, and will remain so until it will be destroyed by some subsequent invasive process [like for instance the observation of the $(u,\sigma)$-position, with $\sigma\neq\rho$, in the quantum machine example].

In fact, also \emph{burnability} is an intrinsic property of a wooden entity, although in this case we cannot observe it in practical terms, for more than once, as the observational process is destructive. But what is important, for the purpose of the definition of intrinsicness, is not if we can perform a number of times the same observation and always collect the same positive answer (as this is only possible for observational processes involving discovery or creation aspects), but that the affirmative answer would always be certain, regardless of the instant we may choose to make the observation.  

To make our point about intrinsicness more explicit, let us consider a classical body $A$ that moves along a given trajectory in the three-dimensional Euclidean space. We consider the property $P_A(\textbf{x}_0)$ of $A$ of being located at $\textbf{x}_0\in\mathbb{R}^3$ (if $A$ is not point-like, then $x_0$ refers to the position of its center of mass). Is $P_A(\textbf{x}_0)$ an intrinsic property of $A$? 

Even by assuming that the observational process is purely non-invasive, i.e., that we can observe the position of $A$ without disturbing it (we don't use here unconventional observational instruments like sawtooth rulers or sticky elastic bands), the answer is clearly negative. Indeed, if, say, $A$ only reaches point $\textbf{x}_0$ at time $t_0$, then, for all $t\neq t_0$, the observation of property $P_A(\textbf{x}_0)$ would yield a negative answer. But even at time $t_0$ the answer could be negative, if for instance the observer decides to translate in space its measuring apparatus, say to a point $\textbf{c}$, so that in relation to the translated reference frame the position of $A$ at time $t_0$ would become $\textbf{x}_0 -\textbf{c}$, and not $\textbf{x}_0$. For these reasons, $P_A(\textbf{x}_0)$ cannot be considered an intrinsic property of $A$. 

Consider then the property $P_A(\mathbb{R}^3)$ of $A$ of being present somewhere in the three-dimensional Euclidean space (i.e., the property of $A$ of being a \emph{spatial entity}). If we ask the same question as before, the answer is now clearly affirmative: $P_A(\mathbb{R}^3)$ is an \emph{intrinsic} property of $A$, as we can predict with certainty the positions it will occupy at any time (by solving the equations of motion), so that we know with certainty that at any time it will be somewhere in space, this being true independently of the kind of trajectory followed by $A$ and of the reference frame that will be adopted by the observer.

But then, what is the fundamental difference between property $P_A(\textbf{x}_0)$ and property $P_A(\mathbb{R}^3)$, that allows us to decree that the former is an \emph{ephemeral}, non-intrinsic property, that can only be associated to $A$ for a moment, and only in relation to a specific observer, whereas the latter is an \emph{intrinsic} property, that can be permanently attached to it, for all observers? 

The answer is quite evident: $P_A(\textbf{x}_0)$ is a \emph{relational} property, whereas $P_A(\mathbb{R}^3)$ is not. Indeed, the spatial position $\textbf{x}_0$ of $A$ can only be defined in relation to a specific observer, associated to a given reference frame. But without any reference to a \emph{specific} observer, to possess a \emph{specific} spatial position is clearly an undefined, meaningless property. So, when we speak about a specific position of $A$, in reality we are speaking about a specific (spatial) relation between $A$ and the observer's frame of reference. In other terms, strictly speaking $P_A(\textbf{x}_0)$ is not a property of entity $A$ per se, but a property of the \emph{composite entity} formed by the combination of entity $A$ and the observer's reference frame.

Now, a relational property cannot be an intrinsic property, seeing that it is a property associated to a specific \emph{configuration} of the composite system formed by the observer and the observed, and as soon as such a configuration is modified (i.e., the specific relation between the two subsystems is broken), the relational property immediately ceases to be actual. In other terms, we must distinguish between properties that entities can have independently of a specific observer, and therefore can be observed by any observer, from those only expressing an exclusive, contingent relation between the entity and a given observer.  
 
Let us open a parenthesis to emphasize that every observation is, in ultimate analysis, a product observation. In the case of the observation of the position of $A$, before actually performing the observation we clearly have to choose a specific reference frame, among an infinite number of potential reference frames. The procedure is therefore similar to that of a \emph{product observation}, that we have described in Section~\ref{How to conjunctly perform incompatible observations}, the only difference being that now the observer perfectly knows which choice s/he is making, so that, with respect to that choice, the outcome of the observation can be predicted in advance, at least in principle.

But knowing which choice is being made doesn't mean to have control over the procedure that produces such choice. This reflection may appear a little strange, as we are used in physics to only consider situations where the physical system and the measuring apparatus are already given. And this is the reason why we usually so strongly believe in determinism. If everything is given, then there is no known reasons to think that the result of the observational process, if properly conducted, wouldn't be pre-determined, whatever the outcome will be. 

But the assumption that the experimental apparatus is given \emph{a priori} cannot be justified in all circumstances. Indeed, as we have seen, when we observe \emph{meet properties} (and more particularly meet properties associated to incompatible observables),  it is part of the very observational procedure to operate a choice of an unpredictable nature, to select the experiment that has to be conducted. Therefore, the apparatus are not always given \emph{a priori}, and this is the reason why the (deterministic) classical prejudice doesn't hold in general (see also the discussion in Ref.~\cite{Piron3}, page 11).

This said, let us consider once more the observation of the position of $A$, and ask the following question: Is it an observation involving a discovery aspect, or a creation aspect? Seeing that we have assumed that the detection of the object is done without producing any disturbance on it, we would be tempted to answer that the observational process only involves a (non-invasive) discovery aspect. However, we must not forget that position is a relational property, associated to the compound entity formed by entity $A$ plus the reference frame, and that the observational process, to be actually performed, requires the choice of a specific reference frame. 

Now, although we are used to think of frames of reference in abstract terms, in practice a reference frame is made manifest by the actual presence in space of a physical observer, be it a human observer or a human observer's measuring apparatus. Let us think of it as a rigid body $R$, to which we have associated a given coordinate system. Then, choosing an observational frame means choosing a specific state for $R$. And this choice will immediately \emph{create} certain positions rather than others for $A$, in the course of its evolution. This is because, as we explained, the specific spatial positions of $A$ are not intrinsic properties, but relational ones. And if we consider them in relation to $R$, then a position $\textbf{x}$ of $A$ at, say, time $t$, is in reality  an internal property of the composite system $\{A,R\}$: a relational property between its two subsystems $A$ and $R$. 

When we choose a specific state for the observational frame $R$, to carry out the position observation, what we do is to invasively act on the composite system $\{A,R\}$, changing the internal relation between its components. In other terms, we deform the system $\{A,R\}$, and obviously such a deformation cannot be considered as a mere discovery. 

So, although we are not so much used to think in this way, to observe the position of a classical object is in fact a process that involves a creation aspect, as the entity's position is literally created during its observation, and this is so even though the chosen detection instrument does not disturb in any way the object in question!

\section{Product observations of relational properties}
\label{Product relational observations}

In the previous section we have considered the possibility to also classify observations by distinguishing those we have called \emph{intrinsic}, which are stably associated to a physical entity, independently of a specific observer, from those we have called \emph{relational}, which are of a much more ephemeral character and can only be attributed to composite systems formed by the physical entity under investigation and a specific observer. 

We have also emphasized that relational properties are literally created during the observation, as they require making a choice about the observational process to be performed (for instance by placing the measuring apparatus in a certain location or state of motion); a choice that will affect the result of the observation. We have done so by considering the simple example of the observation of the position of a classical body, but other examples would have been possible, like for instance the observation of the \emph{velocity} or the \emph{energy} of a classical entity.
   
We want now, in this section, to push further our analysis of relational properties, considering them also in situations where the choice made by the observer is not under her/his control, as in Aert's quantum machine. This will help us getting closer to the understanding of the mystery of quantum measurements.

For this, we consider one of the homogeneous elastic bands used in Aerts' quantum machine. Here we are not thinking of the elastic as an apparatus to be used to measure the position of a point-like particle, as we did in Section~\ref{The (spin) quantum machine}, but as a physical entity per se, about which we want to observe the physical properties. 

An elastic band entity has many interesting properties, like for instance its length when in its unstreched condition, its mass, volume, etc. What we shall consider, however, are some less conventional properties, that we shall call the \emph{left-handedness} and the  \emph{fragmentation} of the elastic (this example is a variant of the ``spaghetti model'' presented in Ref.~\cite{Massimiliano1}). 

\begin{figure}[!ht]
\centering
\includegraphics[scale =.7]{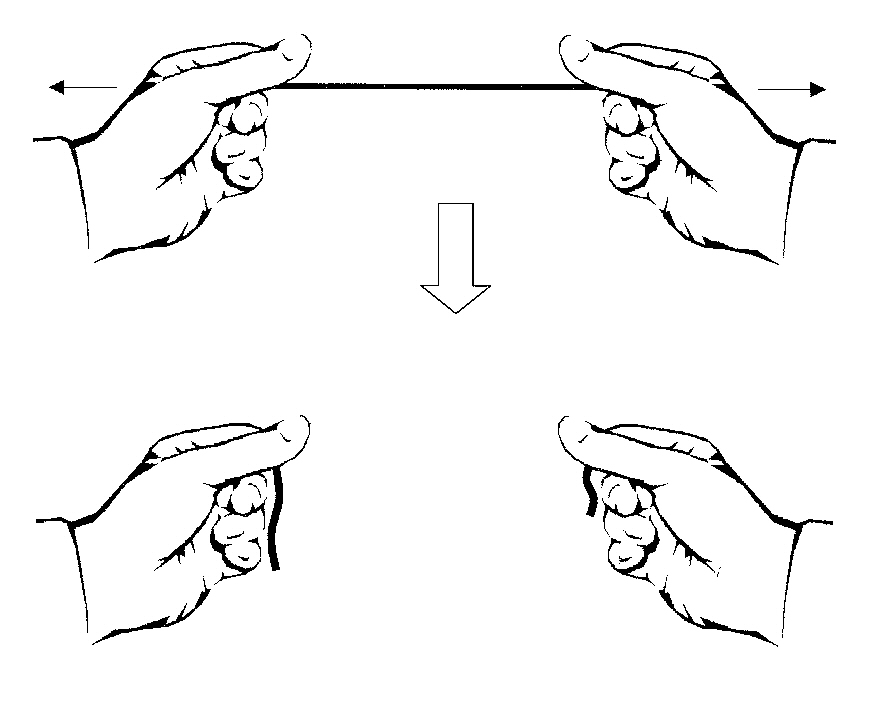}
\caption{Observing the left-handedness of an elastic band.
\label{lefthandedness}}
\end{figure}

To observe the left-handedness of an elastic, the procedure is the following: ``Grab the two ends of the elastic with both hands, then stretch it until it breaks (this one can always do, as an elastic possesses the intrinsic property of \emph{breakability}). If the longest fragment remains in your left hand, then the observation of the left-handedness has been successful (see Fig.~\ref{lefthandedness}).'' 

Now, the observational procedure, to be complete, must specify how to observe the left-handedness also when the elastic is already broken, in two or more fragments (which is always the case once the left-handedness test has been carried out). The prescription is the following: ``If the elastic is already broken, simply do the experiment using the longest fragment.''

Similarly to the position observation of the point-particle in Aerts' quantum machine model, one cannot determine in advance if an elastic is left-handed, and this is not imputable to one's lack of knowledge about its state. Even with a full knowledge of all of the elastic intrinsic properties, down to the molecular level, one cannot predict the outcome of the observation, as the left-handedness property is \emph{created during the observation itself}, according to the specific point where the elastic-entity breaks, as a consequence of fluctuating factors that are not under the control of the observer. 

Imagine now that you have just observed with success the left-handedness of an elastic and want to repeat on that same elastic the same observation. According to the procedure, all you have to do is to take the longest fragment, grab it with your two hands, stretch it until it breaks and observe if the longest fragment remains again in your left hand. 

Clearly, the fact that the previous observation was successful doesn't help at all in predicting the outcome of the second observation. Indeed, to be re-observed, the left-handedness property has to be re-created, and the creation mechanism is non-deterministic. The situation is clearly different compared to the observation of the $(u,\rho)$-position of the point-particle in Aerts' quantum machine. Indeed, as the reader will remember, once the particle's $(u,\rho)$-position is observed (created), subsequent observations of the same $(u,\rho)$-position will produce the same outcome, which means that the $(u,\rho)$-position has become an \emph{intrinsic}, stable property, of the (otherwise static) point-particle entity.

The ephemerality of the left-handedness property of the elastic band is very similar to the ephemerality of most properties of microscopic quantum entities, like for instance the position of an electron. Indeed, quantum mechanics tells us that, in general, not only one cannot predict the position of a microscopic entity prior to the measurement, but neither can one do it a finite time following it, however small such a finite time is. And the same is true for the observation of other properties, like \emph{momentum}, which is incompatible with the position observation, as expressed by Heisenberg's uncertainty principle.

Let us turn for a moment to the question of incompatibility, and ask if also the observation of ephemeral properties like left-handedness can exhibit incompatibility, as it is the case for quantum observables. To do so, let us consider the \emph{fragmentation} property that we mentioned above, which we operationally define as follows: ``Put all the elastic's fragments into a box, shake the box well so that all the fragments get mixed. Then, closing your eyes, open the cover of the box and grab the first fragment you touch. Opening again your eyes, check if it is shorter than half the length of the original unbroken elastic. If it is so, the observation of the fragmentation property has been successful.''

Clearly, for an unbroken elastic the fragmentation property cannot be observed. This means that the inverse property of \emph{non-fragmentation}, defined by the same observational procedure as per above, but by replacing the term ``shorter'' with the term ``longer,'' is an intrinsic property of the unbroken elastic entity. But of course, as soon as the elastic band is broken, as a result for instance of an invasive observation of the left-handedness property, the fragmentation property has some chance to be observed. But it cannot be observed with certainty, as is clear from the fact that the observational procedure (which is non-invasive for the elastic) involves a hidden selection mechanism which is not under the control of the observer. So, fragmentation, as left-handedness, is typically an ephemeral property, and not an intrinsic property. 

However, similarly to the (intrinsic) burnability and floatability properties of the piece of wood entity, the (non-intrinsic) left-handedness and fragmentation properties of the elastic band also entertain an incompatibility relation. Indeed, the observation of the left-handedness property considerably increases the probability that the observation of the fragmentation property will be successful, as is clear from the fact that the greater number of fragments, the higher will be the probability that the observer will pick one which is shorter than one half of the original length. In other terms, by performing first the left-handedness observation we influence the outcome of the successive fragmentation test, which is clearly an expression of experimental incompatibility.

So, incompatibility is a general phenomenon. It doesn't only manifest for the observation of intrinsic properties, but also for ephemeral ones. The present example is about an incompatibility relation between an invasive property involving a creation aspect (left-handedness) and a non-invasive property only involving a discovery aspect (fragmentation). However, one can also easily exhibit examples of incompatibility between ephemeral properties which are both invasive, and we refer for this to the spaghetti example recently presented in Ref.~\cite{Massimiliano1}. 

Having highlighted that properties, despite their ephemeral character, can nevertheless entertain incompatibility relations with other ephemeral properties, and that this fact is not a prerogative of quantum microscopic entities, we want now to clarify that their ephemeral character is a consequence of the fact that they are not, \emph{strictu sensu}, properties of the elastic, but relational properties of the composite system made of the elastic plus the measuring apparatus (the hands of the observer in this case). 

This is quite obvious if one looks closely at the true meaning of a property like the left-handedness. Indeed, to be left-handed, an elastic needs... hands! More precisely, it needs the hands of the observer, i.e., the presence of the measuring apparatus and, more precisely, to entertain a very specific relation with it. Indeed, it is only for as long as the longest fragment remains attached to the left hand of the observer that we can say that the elastic possesses in actuality the left-handedness property. However, to quote the last sentence in Ref.~\onlinecite{Aerts1} ``[...] a measuring process is in a certain sense a unification and then again a separation of the measuring apparatus, and the physical system.'' 

Undeniably, it is one of our unconscious assumptions that when we observe something about a given entity, what we're observing is something that pertains to that entity. But in our previous discussion we have seen that this assumption is not necessarily true, in particular because many times our observational processes can literally create what we are observing, and that this doesn't happen only in the microscopic quantum domain. However, we may still believe that once our observation has created the property which is being observed, such property will then stably belong to the entity, i.e., it will become one of its intrinsic properties. But this, again, is not necessarily the case, as many of the properties we observe are in fact \emph{relational properties}, as is evidenced by our analysis of the position measurement in Section~\ref{Intrinsic VS Relational observations}. 

Now, if properties are relational, they can remain actual only for as long as the specific relation that defines them will not be broken, i.e., for as long as the observer and the observed remain, in a sense, united in a very particular way. But, as noted by Aerts in the above quote, a measure typically involves an initial phase of unification followed by a final phase of separation. This means that a relational property will in general first be created and then destroyed at the end of a measuring process. 

This is what we can clearly see in our paradigmatic example of the observation of the left-handedness property. There is a moment during which the hands of the observer interact with the elastic entity, until a specific outcome is produced. If the outcome is successful, then the left-handedness relational property becomes manifest, i.e., actual, as is evidenced by the fact that the longest fragment dangles from the observer's left hand (see Fig.\ref{lefthandedness}). But this is only true for as long as such a configuration will be maintained, because in the very moment the observer will let go the elastic, the relational left-handedness property will be lost, and the only way to observe it again, i.e., to create it again, is to follow once more the corresponding non-deterministic observational procedure.   

Of course, once the procedure has been completed, and the left-handedness property observed, traces of the interaction that took place could be maintained in the structure of the measuring apparatus, even after the latter has separated from the entity. This we can easily do for instance in our example by adding a camera to the experiment, to register the result of each individual observation. But these traces, like the spots we see on recording screens when we ``detect'' (in fact create) the positions of microscopic entities, are just what they are, traces! That is, memories of ephemeral relational properties that have already ceased to be. 

We will now conclude the present section by providing a general conceptual characterization of quantum measurements. A first important characteristic of these measurements is that they are non-deterministic, in the sense that their outcome can only be predicted in probabilistic terms. As shown by Aerts in his hidden measurement approach~\cite{Aerts4,Aerts7}, the probabilistic aspect of quantum measurements can be described in terms of (hidden) product observations, involving a symmetry breaking mechanism that is beyond the control power  of the observer. Another important characteristic is that quantum measurements involve a pure creation aspect (a fact that is usually described in the literature by the more vague concept of \emph{contextuality}, expressing the fact that the results of our experiments depends in general upon how they are performed). 

Indeed, when we measure the position of an electron, we are not in fact discovering the location where the electron was, prior to our observation, but we are literally creating a spatial localization for it. This is not only because, as we observed in Section~\ref{Intrinsic VS Relational observations}, even the observation of the position of a classical body involves an act of creation, but because, more surprisingly, microscopic quantum entities are truly \emph{non-spatial} entities. By this we mean that, contrary to a classical macroscopic entity, $P_A(\mathbb{R}^3)$ is not an actual property if $A$ is microscopic, since the uncertainty principle doesn't allow us to predict, \emph{not even in principle}, its spatial locations, meaning that these locations cannot pre-exist the observational process (a precise version of this argument can be found in Ref.~\cite{Massimiliano1}). 

Another paradigmatic example, indicating that quantum properties are created by the measurement process, is provided by the famous Kochen-Specker theorem~\cite{Kochen}, proving that it is impossible to attribute specific a priori values to the square of the spin of a spin-$1$ entity, along a given number of axis, in a way that would be compatible with the possible experimental outcomes. But if the entity cannot possess, in general, specific spin values prior to their observation, this is just another way of saying that these values are created by the observational process.  

A last important characteristic of quantum measurements is that they correspond to the observation of ephemeral properties, i.e., of properties that in general can only be actual for a moment. Following our previous analysis, ephemerality appears to be a consequence of the relational character of quantum properties. In other words, if quantum properties are only ephemerically brought into existence by the very observational process that define them, this is because they essentially express the actuality of a relation between the entity and the measuring apparatus; a relation that ceases to be actual as soon as the observation has been performed and the entity and apparatus are again (experimentally) separated. 

Summarizing all this, we can say that a typical quantum measurement can be understood as an observation about a \emph{relational property} between the observed quantum entity and the observing apparatus, entailing a \emph{symmetry breaking} mechanism which is not under the control of the experimenter; a mechanism which selects, in a non-predictable way, a single, possibly deterministic, invasive observational process, of the \emph{creation} kind.

\section{Intermediary observations}
\label{Intermediary observations}

We would like to conclude now our review of the different observational processes, by further explaining the meaning of intermediary observations, which are neither purely classical, nor purely quantum, but truly intermediary -- quantum-like -- observations. We have already mentioned them in Section~\ref{The (spin) quantum machine}, in relation to Aerts' quantum machine, when referring to its $\epsilon$-model variant, where the elastics are not any more uniformly breakable. In this model, $\epsilon$ is a continuous parameter that can be varied from $0$ to $1$. In the $\epsilon = 0$ limit, the elastics can be broken only in a given point, so that the observational processes are purely classical, in the sense that the outcomes can be predicted by only knowing the state of the entity. On the other hand, in the $\epsilon = 1$ limit, the elastics are perfectly uniform and the observational processes, as we explained, are purely quantum, i.e., structurally equivalent to the spin of a spin-$1/2$ quantum entity, reproducing the same transition probabilities of a typical Stern-Gerlach experiment. But in the intermediary situation, $0<\epsilon <1$, the elastics used have a more complex structure, as they are uniformly breakable only in their middle segment (the length of which is proportional to $\epsilon$), so that the outcome of the observational processes can either be predictable, or unpredictable, according to the initial state of the entity (which determines where the point-particle will stick on the elastic, when falling orthogonally onto it). An intermediary situation of this sort gives rise to a state-space structure  that cannot be modelized by a classical phase-space or a quantum Hilbert space, as was demonstrated in~\cite{Aerts3}. In other terms, intermediary observations are associated to more general and complex structures than those originating from purely classical and quantum observations. 

Another interesting example of intermediary observational processes was recently presented by the present author in Ref.~\cite{Massimiliano2}, where a macroscopic quantum machine, called the \emph{$\delta$-quantum-machine}, was introduced, which is able to reproduce the transmission and reflection probabilities of a one-dimensional quantum scattering process by a Dirac delta-function potential. By varying the observational procedures, the machine can exhibit a whole range of intermediary behaviors that cannot be described by a classical or quantum scattering system. 

Here we want to provide a truly elementary example of an intermediary observation that, depending on the state of the system, can give rise to either predictable or unpredictable outcomes for the observational process. In fact, we have already introduced such an example in Section~\ref{Product relational observations}, when we have defined the observation of the \emph{fragmentation} property of an elastic band. According to the state of the elastic band, such an observation can be either totally predictable or perfectly unpredictable. 

To see this, we remark that subsequent observations of the left-handedness (or breakability) property will put the elastic in different states, characterized by an increasing number of its fragments. In general terms, as we already observed, this will increase the probability of observing the fragmentation property. However, as soon as the elastic is in a state such that the length of all its fragments is lesser than one-half of the total original length, then the fragmentation property can be observed with certainty. In other terms, depending on the state of the entity, the outcome can either be predictable or unpredictable, i.e., both situations can occur, for a same system. This is quite different from typical classical observations, where independently of the state of the system the outcome is always predictable, or typical quantum observations, where the outcome is never predictable (if not in probabilistic terms).

To the last statement one can of course object that there are circumstances, also for quantum entities, when certain observations can be predicted in advance, as is the case for example for the electric charge of an electron. This is correct, and is the reason why in general we should not speak of quantum or classical \emph{systems} (or \emph{entities}), but of quantum, classical and intermediary \emph{observational processes} (or \emph{properties}), as a given system can exhibit both quantum and classical properties. 

The crucial point about intermediary observations, which distinguish them from classical and quantum ones, is the fact that, for a same observational process (i.e., a same property) the outcome can be either perfectly predictable, or perfectly unpredictable (but nevertheless possibly predictable in probabilistic terms), according to the state in which the observed system is prepared, and this is a situation that cannot be duly modelized in the ambit of the mathematical structure of a phase space or Hilbert space~\cite{Aerts3}.

\section{Non-spatiality}
\label{Non-spatiality}

It is now time to draw some conclusions from our overview of the observational processes in physical systems. We have analyzed observation taking into consideration different perspectives. We have started by considering that observations can be perfectly \emph{non-invasive}, when they are performed without a specific scope, but can easily become partially, or totally \emph{invasive}, when they address very specific questions to the entities under consideration. This invasiveness of the observational processes can either manifest in terms of \emph{destruction} of the observed properties, or \emph{creation} of them, with modalities that can vary according to the operational definition chosen to define the property which is being observed. We have seen that all this is independent of the fact that the observed systems are macroscopic or microscopic, so that much of the strangeness of quantum observations, in particular their \emph{contextuality}, is in fact also present in conventional entities (and therefore is not so strange after all). 

Another interesting aspect we have emphasized is that \emph{incompatibility} is also a general feature of macroscopic entities, related to the fact that observing an entity will in general affect its state, and therefore also affect the outcome of subsequent observations. This however doesn't mean that we cannot conjunctly observe incompatible observables. To do so, we only have to use a very special observational procedure, called \emph{product observation}. Such a procedure involves a \emph{non-deterministic} act of choice, through which a specific, deterministic observational process is selected among a number of potential ones, and then implemented. 

If such a selection is the result of a \emph{symmetry breaking} mechanism, which remains hidden from the ``view'' of the observer, and is therefore beyond her/his control power, a quantum probabilistic structure may emerge, as it has been shown by Aerts, in his \emph{hidden measurement approach}~\cite{Aerts4,Aerts7}. 

We have also seen that a typical feature of quantum observations (measurements) is the ephemerality of the observed properties, and we have proposed an explanation of this fact in terms of \emph{relational properties}: quantum observations are about relational properties between the system and the measuring apparatus, and since, at the end of a measurement process, their connection is generally severed, this is the reason why the observed properties, once they have been created by the interaction, they are also immediately after destroyed (as soon as that same interaction between the entity and the measuring apparatus is ``switched off.'') This would explain the typical ephemerality (i.e., the non-intrinsicness, or lack of objectification) of quantum properties. 

In our analysis we have also emphasized that relational properties are ubiquitous also in classical physics, as also the observation of position, momentum, energy, etc, of a classical body are in fact the expression of irreducible relational properties between the observed entity and the reference frame associated to the measuring apparatus. Moreover, we have seen that product observations are also ``hidden'' in classical physics, although they are not, as a rule, recognized as such, seeing that the process through which an observer chooses a specific observational process, by choosing for instance  a specific reference frame, is not considered as an integral part of the experiment.   

But if we enlarge our perspective, and duly understand the observation of, for example, the position of (the center of mass of) a macroscopic body $A$, as the observation of a relational property between $A$ and a measuring apparatus $R$, then choosing a specific state for $R$ (here understood in its function of reference frame) is about performing a specific (invasive) act of creation on the composite system $\{A,R\}$. And in that sense, we must acknowledge that classical measurements and quantum measurements are in fact much more similar than what we would have expected.

Using different reference frames to observe the properties of a given body $A$ is therefore not a process as passive as is usually understood. Indeed, since most of the observed properties are irreducibly relational, when we consider the point of view of different observers, associated to different frames of reference, we have to admit that they are not, strictly speaking, observing the same property. In fact, when we speak about the position of $A$, what we truly mean is the ``position-relational property between $A$ and $R$'' -- let us denote it $q_{A,R}$ -- which is a property not of $A$, but of the composite entity $\{A,R\}$. So, when we change the observer, shifting from reference frame $R$ to, say, reference frame $R'$, the $R'$-observer is now observing a different relational property -- $q_{A,R'}$ -- associated to a different composite entity $\{A,R'\}$. 

Of course, the $R'$-observer may also be interested in observing the relational properties associated to the composite system $\{R,R'\}$, and ask if the knowledge of them would allow her/him to deduce the properties of $\{A,R\}$ from those of $\{A,R'\}$, and vice versa. This possibility is made manifest by the existence of certain mathematical \emph{transformations}, the nature of which depends of course on the type of relational properties existing between $R$ and $R'$. When $R$ and $R'$ are for instance two inertial frames, and if we are only interested in time and position relational properties, then these transformations are the well-known \emph{Galilean} or \emph{Lorentz transformations}.

An interesting aspect is that there are certain classes of actions that one can perform on both subsystems $A$ and $R$, such that they  will leave the properties of the composite system $\{A,R\}$ totally unaffected. As it was lucidly pointed out by Poincar\'e beginning of last century~\cite{Poinc}, it is the possibility of these \emph{correlative} actions (called displacements), that leave the properties of the composite system $\{A,R\}$ unchanged, that has brought us human beings, in the course of our evolution, to construct the geometrical Euclidean space, as a convenient way to represent the ensemble of all possible displacement-actions. This of course doesn't mean that our Euclidean space would just be a pure human invention. On that purpose, it is appropriate to quote Aerts, commenting on Poincar\'e's findings~\cite{Aerts2}: 

``[Poincar\'e] analyses how the reality of three dimensional Euclidean (or non-Euclidean) space, has been constructed from our daily experiences as a human being with the objects that are most important for us (rigid bodies), and closely around. This does not mean that this three dimensional
space is an `invention' of humanity. It exists, but the way we have ordered, and later on formalized it, by means of specific mathematical models, does make part of it. In other words, what we call the three dimensional reality of space partly exists in its own and partly exists by the structures that we have constructed, relying on our specific human experience with it.''

In fact, as  Poincar\'e pointed out, also \emph{deformations} of (non-rigid) macroscopic bodies can be understood in terms of displacements, if we conceive a macroscopic body as an aggregate formed by a number of separated sub-entities, so that it is the displacement of some of them, with respect to some others, that would produce the perceived overall deformation effect of the macroscopic body. And this is why also deformations can be visualized, to some extent, in our ordinary space.   

But the construction of our Euclidean spatial theatre, although suitable for representing actions of displacement, certainly cannot represent all possible actions that can be performed on entities. Consider for example a ten dollar bill and the action of tearing it in two pieces~\footnote{We remember this was one of the favorite examples of Constantin Piron, when teaching in Geneva his course of quantum mechanics, whom we had the pleasure of being the assistant about twenty years ago. Another one was his emphasis in distinguishing -- especially when he was writing at the blackboard -- breakable chalks from broken ones!}. Where are the ten dollars once the bill has been torn? Strictly speaking, they disappear from our Euclidean physical space, and in replacement of them, two new ``piece of paper'' entities appear. Clearly, this breaking-action, which is a creation-destruction process, cannot be described by a simple displacement, and this is the reason why after it we cannot any longer represent the ten dollars in our space. 

But we also agree that the ten dollars didn't completely disappear from our reality: they still exist, but in a different sense than before. If before the breaking-action they were ten \emph{actual} dollars, belonging to our physical space, following the breaking-action they have become ten \emph{potential} dollars, not any more belonging to our physical space, but nevertheless still belonging to our reality, as is clear from the fact they can be reassembled, for instance by taping the two pieces together. In other terms, there are different possible \emph{modes of existence} for entities, and when they shift from one mode to the other, this can cause them to leave our ordinary physical space, i.e., not be any more represented inside of it (such a mechanism was recently analyzed in Ref.~\onlinecite{Massimiliano2}, by means of the concept of \emph{process-actuality}, and we refer the interested reader to this paper for further discussion of this topic).  

Now, as we remarked before, when displacement-actions are simultaneously performed upon two subsystems $A$ and $R$, the relational properties among them remain unchanged, and the observation of such correlations is a key ingredient in the construction of our Euclidean space. (To really understand why it is so, we refer the interested reader to Poincar\'e ``Science and Hypothesis'' booklet~\cite{Poinc}, and more particularly to its chapter 4, where the author thoroughly explains how an attentive analysis of the laws governing our sensorial impressions could have brought us to the concept of a geometrical space. In this analysis, a key ingredient is the observation of the existence of certain ``correlative movements,'' that allows to connect certain classes of phenomena, which otherwise we should never have thought of connecting).

On the other hand, as the reader will surely agree, if a breaking-action is simultaneously performed upon $A$ and $R$, all relational properties among them will be radically affected. In other terms, and contrary to a displacement-action, a breaking-action performed on both subsystems $A$ and $R$ will not leave the properties of the composite system $\{A,R\}$ unchanged (in fact, and depending on how the subcomponents $A$ and $R$ have been defined, it can go as far as to completely destroy it). And this is why (following here the same line of reasoning as Poincar\'e) these breaking-actions haven't given rise to the construction of a geometric space that would have played a similar role as the one played by the Euclidean space in relation to displacements.

Now, assuming, as we did, that the specificity of quantum observations (i.e., measurements) is that they correspond to observations of relational properties performed through  acts of creation that involve processes of the breaking-kind (think about the breaking of the elastic band when observing ``its'' left-handedness, and more generally about ``the \emph{actual} breaking the symmetry of the \emph{potential}'' in a ``hidden'' product observation), we can easily understand why quantum entities, like electrons, cannot be conveniently represented within our Euclidean space, as they belong in fact to a larger \emph{space of potentialities}, that we humans have just begun to discover and construct.

The non-ordinary spatiality of quantum entities, which is in fact an expression of their non-locality (non-locality is non-spatiality!) has been thoroughly analyzed by Aerts, in a number of papers~\cite{Aerts2,Aerts3,Aerts4}, and more recently also by the present author~\cite{Massimiliano1,Massimiliano2,Massimiliano3}. For a further analysis of this important concept, we therefore invite the interested reader to consult these works.

It is worth mentioning that the insufficiency of the Euclidean representation also becomes manifest when dealing with relativistic entities, moving at high relative speeds, as space and time (relational) properties get mixed in this case, as shown by Lorentz transformations. To duly represent classical (in the sense of non-quantum) observations, one has then to replace the three-dimensional space by a genuine four-dimensional structure. This however doesn't simply consist in replacing our naive three-dimensional spatial theatre by an equally naive four-dimensional spacetime theatre, in which real change wouldn't be possible, as was evidenced by Aerts in his subtle analysis of the geometric and process-like aspects which are inherent in our construction of reality; an analysis to which we also refer the interested reader~\cite{Aerts4,Aerts15,Aerts16}.

Here we just remark that in our ordinary (non-relativistic) experience with classical macroscopic entities, the length of a rod, or the ticking rate of a clock, are usually considered as typical intrinsic properties, that could be equivalently observed by different inertial observers, independently of their relative state of motion. On the other hand, relativity taught us that this is not generally true: rods' lengths contract and clocks run slower when their velocity relative to an observer is increased.  

These relativistic effects can  of course be understood according to Einsteinian (geometric) interpretation of relativity as generalized parallax effects: the fact that two different inertial observers, $R$ and $R'$, do observe (i.e., measure) different lengths (ticking rates) for a same rod (clock) $A$, simply means that what we thought were intrinsic properties, are in fact observer-dependent relational properties. This being understood, there is no more mystery in the fact that a specific property of composite system $\{A,R\}$, that we improperly call the length (clock rate) of $A$, will generally differ from that same property associated to composite system $\{A,R'\}$. Seeing that $\{A,R\}$ and $\{A,R'\}$ are different systems, there are no a priori reasons to expect they should exhibit identical properties.

On the other hand, these same relativistic effects can also be understood according to a Lorentz (process-like) interpretation of relativity. Indeed, if we understand the composite system $\{A,R'\}$ as the system that is obtained following a certain action of dynamical ``deformation'' on system $\{A,R\}$ -- precisely that action which transforms its subcomponent $R$ into $R'$ -- then we can also advocate that typical relativistic effects are the result of real physical actions performed on the observed entity, and this without the need to invoke the existence of an ether. The point is that, contrary to what is usually assumed, we are not truly observing properties of entity $A$, but properties of a larger composite entity $\{A,R\}$.

It could be objected that relativistic effects, like time dilation, take primarily place not during the period of acceleration that transforms a reference system $R$, say initially at rest with respect to $A$, into a reference system $R'$, moving at a certain constant velocity with respect to it. Therefore, it wouldn't be correct to say that it is the process $R\to R'$ that is responsible for the observed relativistic effects. Well, this depends how we interpret such a statement. Relativistic effects, as we emphasized, are an expression of relational properties which, we do agree, only depend on the relative velocity between $A$ and $R'$, and not on a previous acceleration process that has brought $R'$ to acquire its relative state of motion. However, it is precisely such a prior acceleration that created these velocity-dependent (and not acceleration-dependent) relational properties. In other terms, we must not confuse here the process that created a given property, with the property itself.  

We conclude this work emphasizing that, whether in classical/relativistic or quantum physics, many of our interpretational problems might be the result of an insufficient understanding of the articulated structure of an observational process, and in particular of the fact that many of our observations are in fact, so to say, \emph{meta-observations}, i.e., observations of (intrinsic or relational) properties that are created (or destroyed) by the observational process itself.


\begin{thebibliography}{31}

\bibitem{Piron1} C. Piron, ``Foundations of Quantum Physics,'' W. A. Benjamin Inc., Massachusetts 
(1976).

\bibitem{Piron2} C. Piron, ``La Description d'un Syst\`eme Physique et le Pr\'esuppos\'e de la Th\'eorie Classique,'' Annales de la Fondation Louis de Broglie, \textbf{3}, pp. 131-152 (1978).

\bibitem{Piron3} C. Piron, ``M\'ecanique quantique. Bases et applications,'' Presses polytechniques et universitaires romandes, Lausanne (Second corrected edition 1998), First Edition (1990).

\bibitem{Aerts1} D. Aerts, ``Description of many physical entities without the paradoxes encountered in quantum mechanics,'' Found. Phys., \textbf{12}, pp. 1131--1170 (1982).

\bibitem{Aerts2} D. Aerts, ``An attempt to imagine parts of the reality of the micro-world,'' pp. 3--25, in ``Problems in Quantum Physics II; Gdansk '89,'' eds. Mizerski, J., et al., World Scientific Publishing Company, Singapore (1990). An Italian translation of this article is also available: ``Un tentativo di immaginare parti del micromondo,'' AutoRicerca, Volume 2, pp. 77--109 (2011).

\bibitem{Aerts3} D. Aerts, ``The entity and modern physics: the creation-discovery view of reality,'' in ``Interpreting Bodies: Classical and Quantum Objects in Modern Physics, '' ed. Castellani, E. Princeton Unversity Press, Princeton (1998). 

\bibitem{Aerts4} D. Aerts, ``The Stuff the World is Made of: Physics and Reality,'' pp. 129--183, in ``The White Book of `Einstein Meets Magritte','' Edited by Diederik Aerts, Jan Broekaert and Ernest Mathijs, Kluwer Academic Publishers, Dordrecht, 274 pp. (1999).

\bibitem{Aerts5} D. Aerts, ``The missing element of reality in the description of quantum mechanics of the EPR paradox situation,'' Helv. Phys. Acta, \textbf{57}, pp. 421--428 (1984).

\bibitem{Aerts6} D. Aerts, ``The construction of reality and its influence on the understanding of quantum structures,'' Int. J. Theor. Phys., \textbf{31}, pp. 1815-1837 (1992).

\bibitem{Christiaens} W. Christiaens, ``Some notes on Aerts' interpretation of the EPR-paradox and the violation of Bell-inequalities,''  pp. 259--286, in ``Probing the Structure of Quantum Mechanics: Nonlinearity, Nonlocality, Computation and Axiomatics,'' World Scientific, Singapore, 394 pp. (2002).

\bibitem{Aerts7} D. Aerts, ``A possible Explanation for the Probabilities of Quantum Mechanics,'' J. Math, Phys., \textbf{27}, pp. 202--210 (1992).

\bibitem{JOCS} ``Sheldrake and his critics: the sense of being glared at,'' Journal of Consciousness Studies, \textbf{12}, pp. 1--126 (2005).

\bibitem{Heisenberg} W. Heisenberg, ``The Physical Principles of Quantum Theory,'' University of Chicago Press (1930).

\bibitem{Einstein} A. Einstein, B. Podolsky and N. Rosen, ``Can Quantum-Mechanical Description of Physical Reality Be Considered Complete?,'' Phys. Rev., \textbf{47}, pp. 777-780 (1935).

\bibitem{Massimiliano1} Sassoli de Bianchi, M., ``Ephemeral Properties and the Illusion of Microscopic Particles,'' Foundations of Science, 16, No. 4 pp. 393--409 (2011); doi: 10.1007/s10699-011-9227-x. An Italian translation of the article is also available: ``Propriet\'a effimere e l'illusione delle particelle microscopiche,'' AutoRicerca, Volume 2, pp. 39--76 (2011).

\bibitem{Aerts11} D. Aerts, ``Reality and probability: introducing a new type of probability calculus,'' pp. 205--229, in ``Probing the Structure of Quantum Mechanics: Nonlinearity, Nonlocality, Computation and Axiomatics,'' World Scientific, Singapore, 394 pp. (2002).

\bibitem{Aerts12} D. Aerts, ``Being and change: foundations of a realistic operational formalism,'' pp. 71--110, in ``Probing the Structure of Quantum Mechanics: Nonlinearity, Nonlocality, Computation and Axiomatics,'' World Scientific, Singapore, 394 pp. (2002).

\bibitem{Aerts13} D. Aerts, ``Quantum mechanics: structures, axioms and paradoxes,'' pp. 141--205, in ``The Indigo Book of `Einstein Meets Magritte','' Edited by Diederik Aerts, Jan Broekaert and Ernest Mathijs, Kluwer Academic Publishers, Dordrecht, 239 pp. (1999).

\bibitem{Smets} S. Smets, ``The modes of physical properties in the logical foundations of physics,'' Logic and Logical Philosophy, \textbf{14}, pp. 37--53 (2005).

\bibitem{Con1} J. H. Conway and S. Kochen, ``The free will theorem,'' Found. of Physics, \textbf{36}, pp. 1441--1473 (2006).

\bibitem{Con2} J. H. Conway and S. Kochen, ``The strong free will theorem,'' Notices of the American Mathematical Society, \textbf{56}, pp. 226--232 (2009).

\bibitem{Aerts-3d} D. Aerts, B. Coecke, B. D'Hooghe and F. Valckenborgh, A mechanistic macroscopic physical entity with a three-dimensional Hilbert space description, Helv. Phys. Acta, 70, 793--802 (1997).

\bibitem{Aerts14} D. Aerts, ``Quantum Structures, Separated Physical Entities and Probability, '' Found. of Physics, \textbf{24}, p. 1227 (1994).

\bibitem{Coecke1} B. Coecke, ``Hidden measurement representation for quantum entities described by finite dimensional complex Hilbert spaces,'' Found. Phys., \textbf{25}, p. 203 (1995).

\bibitem{Coecke2} B. Coecke, ``Generalization of the proof on the existence of hidden measurements to experiments with an infinite set of outcomes,'' Found. Phys. Lett., \textbf{8}, p. 437 (1995).

\bibitem{Coecke3} B. Coecke, ``New examples of hidden measurement systems and outline of a general scheme,'' Tatra Mountains Mathematical Publications, \textbf{10}, p. 203 (1996).

\bibitem{Massimiliano2} Sassoli de Bianchi, M., ``The $\delta$-quantum machine, the $k$-model, and the non-ordinary spatiality of quantum entities,'' To appear in: Foundations of Science, arXiv:1104.4738v2 [quant-ph] (2011).

\bibitem{Kochen} S. Kochen and E.P. Specker, ``The problem of hidden variables in quantum mechanics,'' Journal of Mathematics and Mechanics, \textbf{17}, pp. 59--87 (1967).

\bibitem{Gleason} A.M. Gleason, Measures on the Closed Subspaces of a Hilbert Space, J. Math. Mech., 6, 885--893 (1957).

\bibitem{Poinc} H. Poincar\'e, ``La science et l'hypoth\`ese,'' Flammarion, Paris (1902).

\bibitem{Massimiliano3} M. Sassoli de Bianchi, ``From permanence to total availability: a quantum conceptual upgrade,'' To appear in: Foundations of Science, doi: 10.1007/s10699-011-9233-z.

\bibitem{Aerts15} Aerts, D., ``Relativity theory: what is reality?'' Found. Phys. \textbf{26}, pp. 1627--1644 (1996).

\bibitem{Aerts16} Aerts, D., ``Towards a framework for possible unification of quantum and relativity theories,'' Int. J. Theor. Phys. \textbf{35}, pp. 2399--2416 (1996).


\end{thebibliography}
\end{document}